\documentclass{aa}  

\usepackage{graphicx}
\usepackage{txfonts}
\usepackage{booktabs}
\newcommand{\Vr}{\vec{r}}
\newcommand{\ud}{{\rm d}}

\newcommand{\bperp}{B_\perp}
\newcommand{\bperpm}{$\langle B_\perp \rangle$}
\newcommand{\betaobs}{$\beta_{\rm obs}$}
\newcommand{\betasyn}{$\beta_{\rm syn}$}
\newcommand{\bperpT}{$\tilde{B}_{\perp}$}

\defcitealias{Padovani2021}{P21}
\defcitealias{Guzman2011}{G11}
\usepackage[breaklinks, colorlinks, citecolor=blue]{hyperref}
\begin{document}

   \title{A new analytical model of the cosmic-ray energy flux for Galactic diffuse radio emission}

    \authorrunning{Bracco et al.}

   \author{Andrea Bracco\inst{1,2}
          \and
          Marco Padovani\inst{1}
          \and 
          Daniele Galli\inst{1}
          }

   \institute{INAF – Osservatorio Astrofisico di Arcetri, Largo E. Fermi 5, 50125 Firenze, Italy\\
   \email{andrea.bracco@inaf.it}
   \and
   Laboratoire de Physique de l'Ecole Normale Sup\'erieure, ENS, Universit\'e PSL, CNRS, Sorbonne Universit\'e, Universit\'e de Paris, F-75005 Paris, France \\
             }

   \date{Received 15 February 2024; accepted 29 February 2024}
 
  \abstract
   {
Low-frequency radio observations of diffuse synchrotron radiation offer a unique vantage point for investigating the intricate relationship between gas and magnetic fields in the formation of structures within the Galaxy, spanning from the diffuse interstellar medium (ISM) to star-forming regions.

Achieving this pivotal objective hinges on a comprehensive understanding of cosmic-ray properties, which dictate the effective energy distribution of relativistic electrons, primarily responsible for the observable synchrotron radiation. 
Notably, cosmic-ray electrons (CRe) with energies between  100 MeV and 10 GeV play a crucial role in determining the majority of the sky brightness below the GHz range. However, their energy flux ($j_e$) remains elusive due to solar modulation.

We propose deriving observational constraints on this energy gap of interstellar CRe through the brightness temperature spectral index of low-frequency radio emission, here denoted as \betaobs. We introduce a new parametric analytical model that fits available data of $j_e$ in accordance with the \betaobs\ values measured in the literature between 50 MHz to 1 GHz for diffuse emission in the Milky Way. Our model allows to account for multiple observations considering magnetic-field strengths consistent with existing measurements below 10 $\mu$G. We present a first all-sky map of the average component of the magnetic field perpendicular to the line of sight and validate our methodology against state-of-the art numerical simulations of the diffuse ISM.

This research makes headway in  modeling Galactic diffuse emission with a practical parametric form. It provides essential insights in preparation for the imminent arrival of the Square Kilometre Array.
   
   }


   \maketitle
%
\section{Introduction}\label{sec:intro}

Low-frequency radio emission observations are paving the way for a comprehensive study of Galactic magnetic fields traced by synchrotron radiation. The breakthrough in sensitivity of current and upcoming radio telescopes, including the Low Frequency Array \citep[LOFAR,][]{vanHaarlem2013}, the Long Wavelength Array \citep[LWA,][]{Dowell2017}, the New Extension in Nançay Upgrading LOFAR \citep[NenuFAR,][]{Zarka2012}, the C-Band All-Sky Survey \citep{Jones2018}, and the Square Kilometre Array \citep[SKA,][]{Dewdney2009} along with its precursors, promises the most detailed multi-scale description of the Galactic magnetic field. This description encompasses both its topology and its strength, ranging from the diffuse interstellar medium (ISM) to star-forming regions \citep{Heald2020}.

Achieving this paramount objective, however, relies on the thorough characterization of cosmic-ray (CR) properties, acceleration, and propagation. These factors determine the effective energy distribution of relativistic electrons, which are primarily responsible for the observable synchrotron radiation\footnote{Positrons and secondary electrons also contribute to synchrotron radiation but at a few \% level \citep[][]{Orlando18,Ponnada2024}} \citep[e.g.,][]{GinzburgSyrovatskii1964, PadovaniGalli2018}. 
In particular, low-energy cosmic-ray electrons (CRe) between 100 MeV and 10 GeV are those relevant for most of the radio emission detected below the GHz range \citep[][hereafter \citetalias{Padovani2021}]{Padovani2021}. Unfortunately, because of solar modulation, the energy flux of these electrons ($j_e$) cannot be measured from near-earth direct observations \citep[e.g.,][]{Gabici2022}. Hence, $j_e$ is usually interpolated in the GeV window \citep[e.g.,][]{Orlando18,PadovaniGalli2018,Padovani2018,Unger2023,Bracco2023} between the MeV range observed with the Voyager spacecrafts \citep{Cummings2016,Stone2019} and the hundred-of-GeV range measured by facilities including Fermi LAT \citep{Ackermann2010}, Pamela \citep{Adriani2011}, and the Alpha Magnetic Spectrometer \citep[AMS,][]{Aguilar2014}.

Building upon the pioneering works of \citet{Rockstroh1978},  \citet{Strong1978}, and \citet{Strong2000}, and following up the investigation conducted by \citetalias{Padovani2021}, in this paper we propose that observational constraints on the missing energy window of interstellar CRe can be obtained through the spectral index of low-frequency radio emission, denoted as \betaobs. We introduce a new analytical-parametric model that accurately fits $j_e$, considering the values of \betaobs\ measured in the  literature between 45 MHz and 408 MHz for Galactic diffuse emission.  
We use our models of $j_e$ to estimate the strength of the Galactic magnetic field averaged along the line of sight. Our observational results are discussed through a comparison with synthetic data generated from state-of-the-art magneto-hydrodynamic (MHD) simulations of the diffuse ISM.  

By incorporating the constraint of \betaobs\ in the shape of $j_e$, we demonstrate an enhanced ability to statistically account for numerous observations between 50 MHz and 1 GHz. This improvement is achieved without resorting to unrealistic values for the strength of the Galactic magnetic field, which, as measured through Zeeman splitting in the diffuse ISM and rotation measures of pulsars, averages less than 10 $\mu$G \citep{Heiles2005, Sobey2019}. This research represents a notable advancement in the modeling of Galactic diffuse emission, offering crucial insights to prepare for the arrival of the SKA.

The paper is organized as follows: in Sect.~\ref{sec:beta}, we introduce the basic formalism of synchrotron emission and review several observations of \betaobs\ at low frequencies reported in the literature; in Sect.~\ref{sec:je}, we present our $j_e$ model; in Sect.~\ref{sec:B}, we derive an all-sky map of the magnetic-field strength and validate our methodology with MHD simulations. The paper concludes with a discussion (Sect.~\ref{sec:discussion}) and a summary (Sect.~\ref{sec:sum}).

\section{Observations of the synchrotron spectral index}\label{sec:beta}

In this section, we first introduce the formalism linking $j_e$ and \betaobs\ (Sect.~\ref{ssec:emis}) and then summarize the observed values of \betaobs\ below 1~GHz (Sect.~\ref{ssec:beta}). We provide a list of observational results that does not aim to be comprehensive but rather representative of what can be found in the literature.

\subsection{From first principles to brightness temperature}\label{ssec:emis}

Galactic synchrotron diffuse emission arises from the interaction of interstellar magnetic fields with CRe \citep[e.g.,][]{GinzburgSyrovatskii1964, Rybicki1979}. These relativistic particles gyrate about the field lines emitting non-thermal radiation dependent on their energy ($E$) and the magnetic-field strength perpendicular to the sightline (${B}_{\perp} = |\vec{B}_{\perp}|$). The total synchrotron emissivity in units of power per unit volume, frequency ($\nu$),
and solid angle is the sum of the emissivities linearly polarized along ($\varepsilon_{\nu,\parallel}$) and across ($\varepsilon_{\nu,\perp}$) $\vec{B}_{\perp}$:

\begin{eqnarray}\label{eq:epsnu}
\varepsilon_{\nu,\|}(\Vr) &=& \int_{m_{e}c^{2}}^{\infty}\frac{j_{e}(E)}{\varv_{e}}P_{\nu,\|}^{\rm em}(E,{\bperp}(\Vr))\,\ud E\\\nonumber
\varepsilon_{\nu,\perp}(\Vr) &=& \int_{m_{e}c^{2}}^{\infty}\frac{j_{e}(E)}{\varv_{e}}P_{\nu,\perp}^{\rm em}(E,{\bperp}(\Vr))\,\ud E,
\end{eqnarray}
where $\varv_e$ is the CRe velocity, $m_e$ is their mass, $c$ is the speed of light, $\Vr$ is the position, and $P^{\rm em}_{\nu, \perp~\rm{\rm or}~\parallel}$ is the power per unit frequency for the two polarizations (see also equation 2 in \citetalias{Padovani2021} and \citealt{Longair2011}). 
By integrating Eq.~\ref{eq:epsnu} along the line of sight (LOS), and accounting for the telescope beam ($\Theta_{\rm T}$), one obtains the synchrotron brightness temperature as 

\begin{equation}\label{eq:tb}
    \frac{T_{\rm b} (\nu)}{[{\rm K}]} = \frac{2\ln{(2)}~c^2 10^{-23}}{\pi k_{\rm B} \theta_{\rm T}^2 \nu^2} \left[ \Theta_{\rm T} \circledast \int_{\rm LOS} (\varepsilon_{\nu,\|} + \varepsilon_{\nu,\perp})\ud r \right] ,
\end{equation}
where $k_{\rm B}$ is the Boltzmann constant and the $\circledast$-symbol represents the convolution with a Gaussian-approximated $\Theta_{\rm T}$ given its full width half maximum (FWHM), $\theta_{\rm T}$, in units of radians. Observationally, 
$T_{\rm b} (\nu)$ mostly results in a power-law function of $\nu$ with spectral index \betaobs\ \citep[e.g.,][]{ReichReich88a,ReichReich88b,Roger+1999}.
From Eqs.~\ref{eq:epsnu} and \ref{eq:tb}, and as already shown by \citetalias{Padovani2021}, the value of \betaobs\ is not expected to be constant but rather to vary depending on the MHD properties of the intervening ISM and on $j_e$. {Notably, it has long been known that energy losses may contribute to modifying the synchrotron spectral index, as illustrated by several observations at low frequencies in the radio band \citep[e.g.,][]{Bridle1967,Sironi1974,Webster1975} as well as by diffusion and convection models of CRe propagation \citep[e.g.,][]{Strong1977}. 

In this work we exploit these dependencies and invert the argument using ancillary observations of \betaobs\ to constrain $j_e$ and ${B}_{\perp}$. In the following, we will distinguish the observed values of the spectral index, \betaobs, from those derived from models, denoted as \betasyn.   

\subsection{Observed spectral indices below 1 GHz}\label{ssec:beta}
\begin{figure}[!t]
\begin{center}
\resizebox{1.0\hsize}{!}{\includegraphics{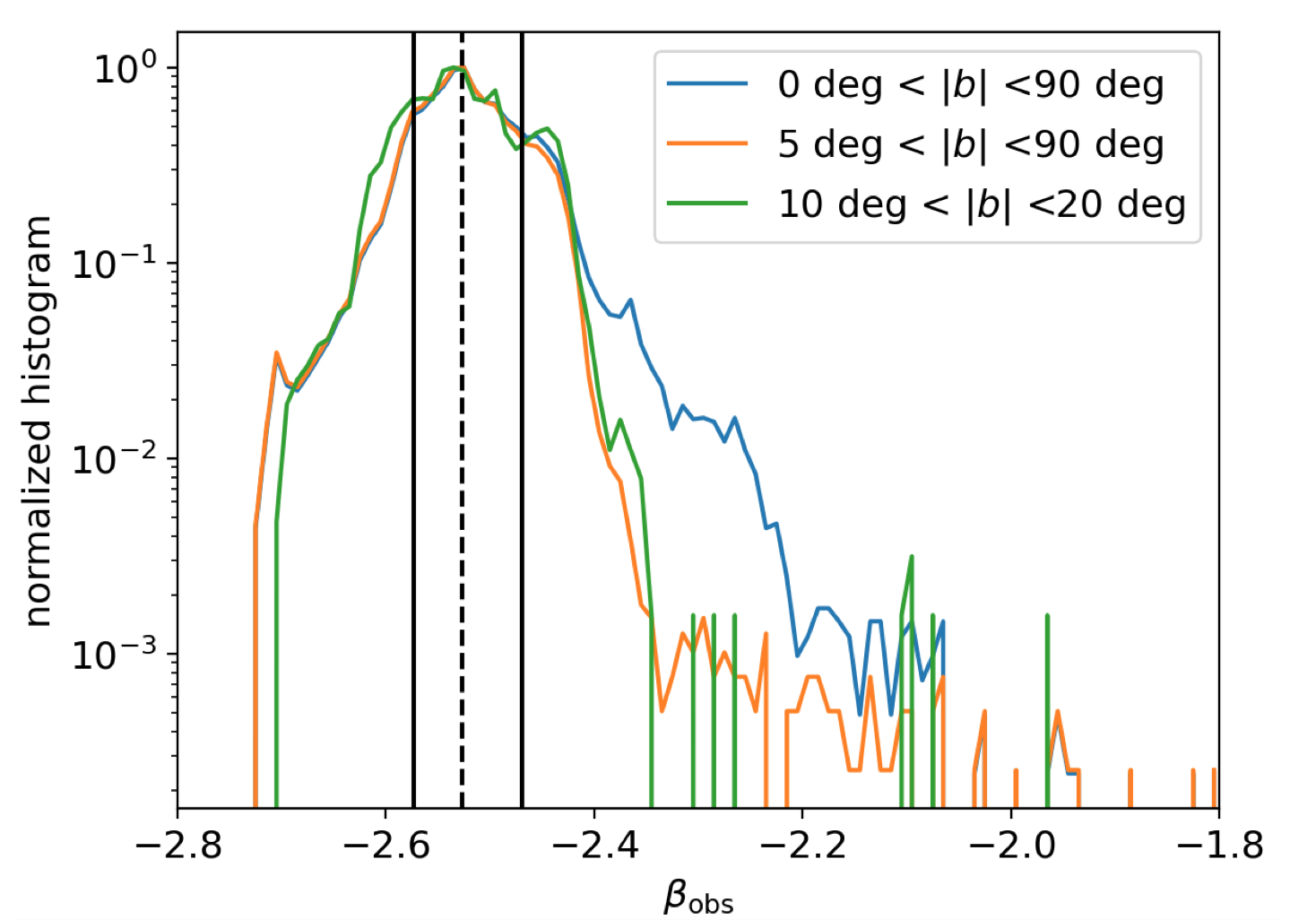}}
\caption{Normalized histograms of the observed spectral indices, \betaobs, between 45 MHz and 408 MHz from \citetalias{Guzman2011} as a function of Galactic latitude. The 50th, the 16th, and the 84th percentiles of the orange histogram are marked in black-dashed and -solid lines, respectively.}
\label{fig:histoguz}
\end{center}
\end{figure}
\begin{figure*}[!h]
\begin{center}
\resizebox{1.0\hsize}{!}{\includegraphics{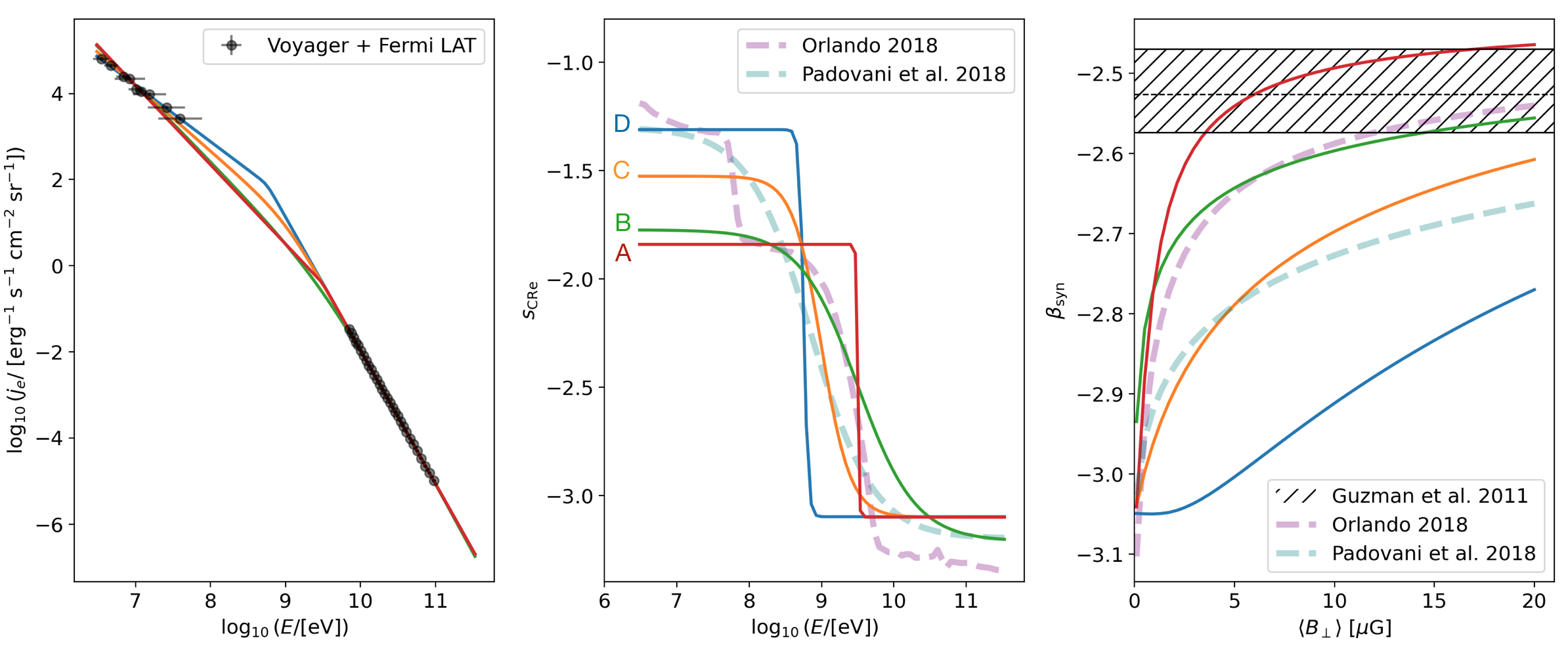}}
\caption{Our models for $s_{\rm CRe}$ obtained with Eq.~\ref{eq:mod} and their impact on the modelled synchrotron spectral index $\beta_{\rm syn}$. Colours correspond to the same models in all panels. {\it Left panel}: $j_e$ from models A, B, C, and D with parameters listed in Table~\ref{tab:models} that fit the black data points from Voyager \citep[$E<10^{8}$ eV, ][]{Cummings2016,Stone2019} and Fermi LAT \citep[$E>10^{10}$ eV, ][]{Ackermann2010}. {\it Central panel}: $s_{\rm CRe}$ values corresponding to the same models as in the left panel. {\it Right panel}: values of $\beta_{\rm syn}$ for the same models as in the other panels with the observed spectral indices from \citetalias{Guzman2011} in hatches -- the 16th, the 50th, and the 84th percentiles are marked as in Fig.~\ref{fig:histoguz}. Models for $s_{\rm CRe}$ and $\beta_{\rm syn}$ using $j_e$ from \citet{Orlando18} and \citet{Padovani2018} are also shown in dashed-purple and -teal lines in the central and right panels, respectively.}
\label{fig:models}
\end{center}
\end{figure*}
Measurements of the synchrotron spectral index from diffuse emission in the Galaxy generally vary between $-2$ and $-3$ depending on the observed frequency range, the sky coverage, and the FWHM of the instrument. Most \betaobs\ at frequencies below 408 MHz have been recently measured with single dipole antennas designed for low-frequency cosmological experiments, such as the Experiment to Detect the Global EoR Signature \citep[EDGES,][]{Bowman2008} and the Large-Aperture Experiment to Detect the Dark Ages \citep[LEDA,][]{Price2018}. These experiments measured  $-2.59 < \beta_{\rm obs} < -2.54$ between 50 MHz and 100 MHz \citep{Mozdzen2019}, $-2.62 < \beta_{\rm obs} < -2.60$ between 90 MHz and 190 MHz \citep{Mozdzen2017}, and $-2.55 < \beta_{\rm obs} < -2.45$ between 50 MHz and 87 MHz \citep[][]{Spinelli2021}. 
While these measurements integrate the synchrotron flux over the full accessible sky, higher angular resolution observations of \betaobs\ are also available. In particular, in this work we refer to one of the all-sky \betaobs\ maps of reference for low-frequency studies derived by \citet{Guzman2011}, hereafter \citetalias{Guzman2011}, between 45 MHz and 408 MHz using data from \citet{Alvarez1997b}, \citet{Maeda1999}, and \citet{Haslam1982}\footnote{The map can be downloaded at the following link: \url{http://cdsarc.u-strasbg.fr/viz-bin/qcat?J/A+A/525/A138}}. This map, with a FWHM of 5 degrees, shows large variations of \betaobs\ ranging between $-2.7$ and $-2.1$. In Fig.~\ref{fig:histoguz}, we show the normalized histograms of these \betaobs\ as a function of Galactic latitude ($b$). Most of the shallow values of \betaobs\ are confined to the Galactic plane, $|b|<5^{\circ}$ (compare the blue and orange curves), possibly due to free-free absorption of radio photons along the LOS \citep[e.g.,][]{Stanislavsky2023}. However, the distribution of the values of \betaobs\ at intermediate to high Galactic latitude above $|b|>5^{\circ}$ is stable with a rather normal shape (see the green and orange curves). The 50th, 16th, and 84th percentiles of the orange histogram are $-2.53$, $-2.57$, and $-2.47$, respectively. 

More recent all-sky maps at the degree-scale level, depicting \betaobs\ values from 35 MHz to 80 MHz, are available from the LWA \citep[][]{Dowell2017}. These maps exhibit good agreement with those presented by \citetalias{Guzman2011}, albeit displaying slightly shallower \betaobs\ values at intermediate Galactic latitudes. The authors did not extensively investigate these differences, attributing them to ({\it i}\/) variations in frequency coverage and ({\it ii}\/) residual systematic effects present in their data.

Remarkably, \betaobs\ maps of external galaxies, such as NGC 5775 and NGC 891, have been produced at low frequency with LOFAR and the Very Large Array \citep{Mulcahy2018, Heald2020}. We notice that, in both cases, the maps appear consistent with the \betaobs\ map of \citetalias{Guzman2011} outside of their galactic halo.

At higher frequencies, above 408 MHz, the \betaobs\ values for diffuse Galactic emission typically converge towards $-3$. A recent example of this comes from the analysis of MeerKAT data (with FWHM of approximately 1 degree) between 971 MHz and 1075 MHz at intermediate Galactic latitudes, which yielded an average \betaobs\ value of $-2.76 \pm 0.15$ \citep[and references therein]{Irfan2022}.

\section{Modeling the CRe energy flux}\label{sec:je}


The ability of models to reproduce the observed synchrotron sky brightness and \betaobs\ strongly relies on the choice of $j_e$ and its approximations (see Eq.~\ref{eq:epsnu}). As discussed in \citetalias{Padovani2021}, most theoretical models and numerical simulations of Galactic synchrotron emission assume fixed power laws for the energy distribution of CRe depending on the frequency range \citep[e.g.,][]{Sun2008, Waelkens2009, Sathyanarayana2017, Reissl2019, Wang2020}. However, this assumption may lead to significant biases in the interpretation of synchrotron data, notably concerning the spatial variations of the spectral index both at low and high frequency. The energy dependence of the slope of $j_e$ between 100 MeV and 50 GeV should be taken into account.

Some effort has been done in this direction, as shown for instance in the works by \citet{Padovani2018} and \citet{Orlando18}. The former proposed an analytical four-parameter model that reproduces the power-law behaviour of $j_e$ measured at low and high energies (see Sect.~\ref{sec:intro}). The latter is based on a combination of multi-frequency observations, Voyager measurements, and the CRe propagation code called GALPROP \citep[e.g.][for a recent release]{Porter2022}. However, as shown with dashed lines in the right panel of Fig.~\ref{fig:models}, both cases struggle in reproducing the range of \betaobs\ described in Sect.~\ref{ssec:beta} (see hatches from \citetalias{Guzman2011})\footnote{We note that \citet{Orlando18} developed their $j_e$ model including the radio data employed by \citetalias{Guzman2011}. However, since their post-processing was not the same, in particular concerning their methods of component separation (i.e., thermal and non-thermal components), the respective \betaobs\ are different. We choose to consider the work of \citetalias{Guzman2011}, as it is widely referred to in the radio community at low frequency. It is beyond our scopes to discuss at length the inconsistencies between these two works.} without resorting to unlikely large values for the average $B_{\perp}$ along the LOS (hereafter, $\langle B_{\perp} \rangle$) of the order of 20 $\mu$G. 

In this context, we propose a new analytical model for $j_e$ tailored to capture its variation in the energy gap encompassing 1 GeV. We introduce a four-parameter model that describes the logarithmic derivative of $j_e$ with respect to $E$ ($s_{\rm CRe}$) as follows 

\begin{equation}\label{eq:mod}
s_{\rm CRe}  =  \frac{d \log_{10}{j_e}}{d \log_{10}{E}} = \frac{s_1-s_2}{1+(E/E_0)^{\gamma}} + s_2,
\end{equation}
where $\gamma = 1/\log{(\Delta E /[\rm eV])}$, and $\Delta E$ represents the energy range around $E_0$ at which the transition between the low-energy and high-energy slopes, $s_1$ and $s_2$, occurs. For comparison, the $s_{\rm CRe}$ curves from \citet{Orlando18} and \citet{Padovani2018} are displayed in the central panel of Fig.~\ref{fig:models}.

Instead of fitting the full set of independent data at once (from those of $j_e$ to the radio ones) to illustrate the impact of our parameters, in Fig.~\ref{fig:models}, we show four realizations of our model that correspond to models A to D in Table~\ref{tab:models}. 
\begin{table}[!h]
\begin{center}
\caption{Model parameters entering Eq.~\ref{eq:mod} and shown in Figs.~\ref{fig:models} and \ref{fig:oriona}.}
\begin{tabular}{c c c c c c} 
\toprule\toprule
model & $s_1$ & $s_2$ & $E_0$ & $\gamma$ & region \\ 
 &  &  &  [GeV] & &  \\
\midrule
A & $-1.8$ & $-3.1$ & $3.2$ & $45$ & diffuse ISM\\
B & $-1.8$ & $-3.2$ & $3.2$ & $2.2$ & diffuse ISM \\
C & $-1.5$ & $-3.1$ & $1$ & $1.1$ & diffuse ISM \\
D & $-1.3$ & $-3.1$ & $0.6$ & $14$ & diffuse ISM \\ 
\midrule
E & $-1.2$ & $-3.1$ & $3.2$ & $45$  &  Orion-Taurus ridge\\
\bottomrule
\label{tab:models}
\end{tabular}
\end{center}
\end{table}

From left to right, models are shown with the same color for $j_e$, $s_{\rm CRe}$, and \betasyn, respectively. Despite all fitting the Voyager and Fermi-LAT data in the left panel, they behave differently in the 1-GeV energy window, as depicted in the central panel. This energy range determines the observable brightness and the corresponding \betasyn, as illustrated in the right panel. 
Model A, characterized by a sharp change in slope of $j_e$ at about 3 GeV (not distinguishable from a step function in $s_{\rm CRe}$), is the type of model that converges in both energy and frequency for values of $\langle B_{\perp} \rangle$ below 10 $\mu$G. Our approach is purely phenomenological and, on average, suggests the existence of two power laws of CRe contributing to synchrotron emission. 

\begin{figure}[!t]
\begin{center}
\resizebox{1.0\hsize}{!}{\includegraphics{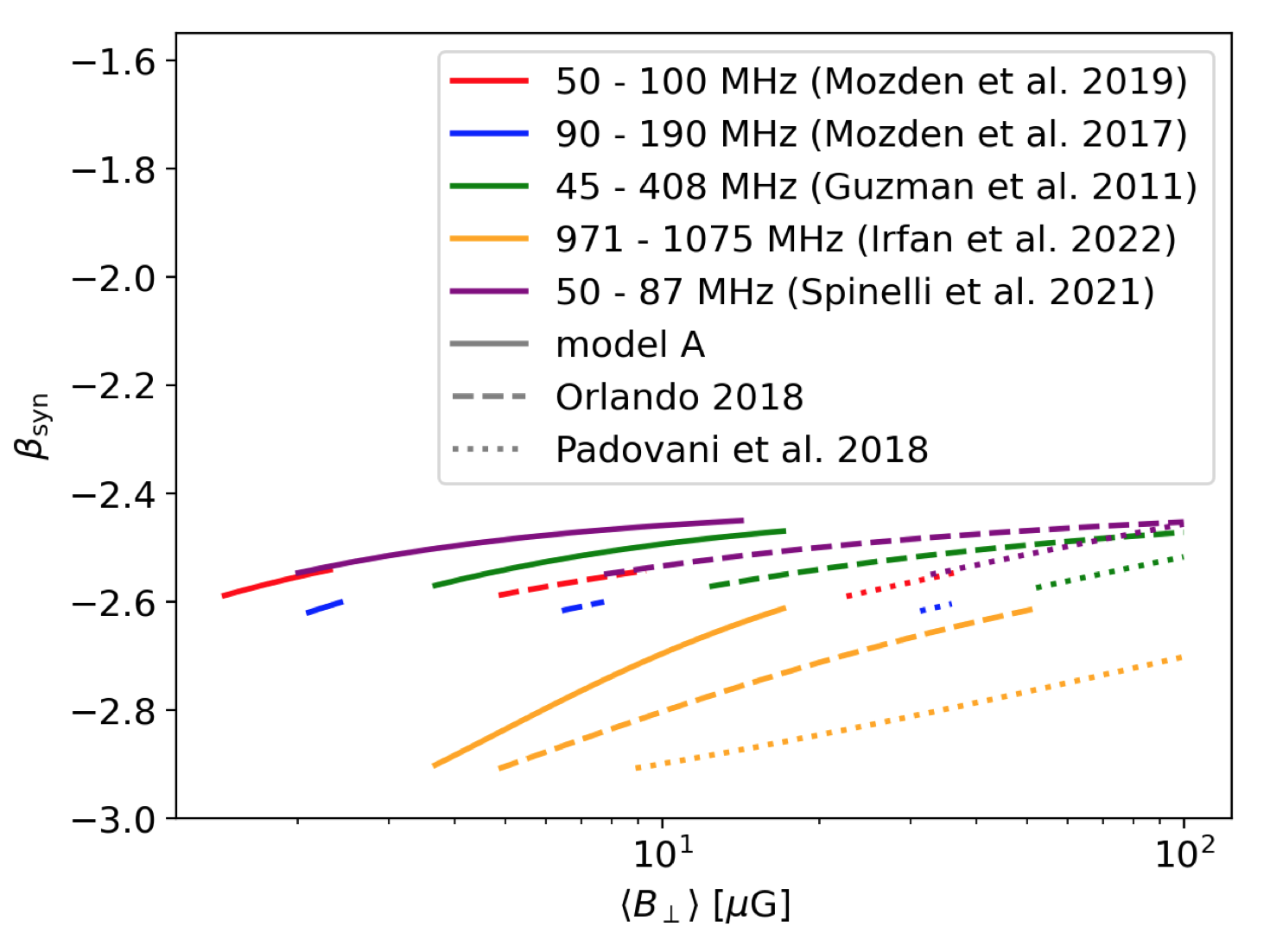}}
\caption{Modeled synchrotron spectral indices, $\beta_{\rm syn}$, reproducing various observational results of $\beta_{\rm obs}$ (in colours) with model A (solid lines), \cite{Orlando18} (dashed lines), and \cite{Padovani2018} (dotted lines). Data are the following: between 50 - 100 MHz (red), and 90 - 190 MHz (blue) using EDGES \citep{Mozdzen2017,Mozdzen2019}; between 45 - 408 MHz using a set of multiple observations (\citetalias{Guzman2011}); between 971 - 1075 MHz using MeerKAT \citep{Irfan2022}; between 50 and 87 MHz using LEDA \citep{Spinelli2021}.}
\label{fig:beta_var}
\end{center}
\end{figure}

Using model A as a reference, in Fig.~\ref{fig:beta_var}, we also compare our estimates of \betasyn\ with the models of \citet{Padovani2018} and \citet{Orlando18} for a number of observations below 1~GHz (see Sect.~\ref{ssec:beta}). As for the range of \betaobs\ from \citetalias{Guzman2011}, we show that model A  better reproduces \betaobs\ both at low and high frequency for realistic values of $\langle B_{\perp} \rangle < 10\, \mu$G. We notice that most single-dipole-antenna experiments produce \betaobs\ values consistent with smaller $\langle B_{\perp} \rangle$ than other instruments. This is possibly due to averaging Galactic diffuse emission and $B_{\perp}$ over larger sky areas.   

\section{Estimate of the magnetic-field strength}\label{sec:B}
\begin{figure}[!t]
\begin{center}
\resizebox{1.0\hsize}{!}{\includegraphics{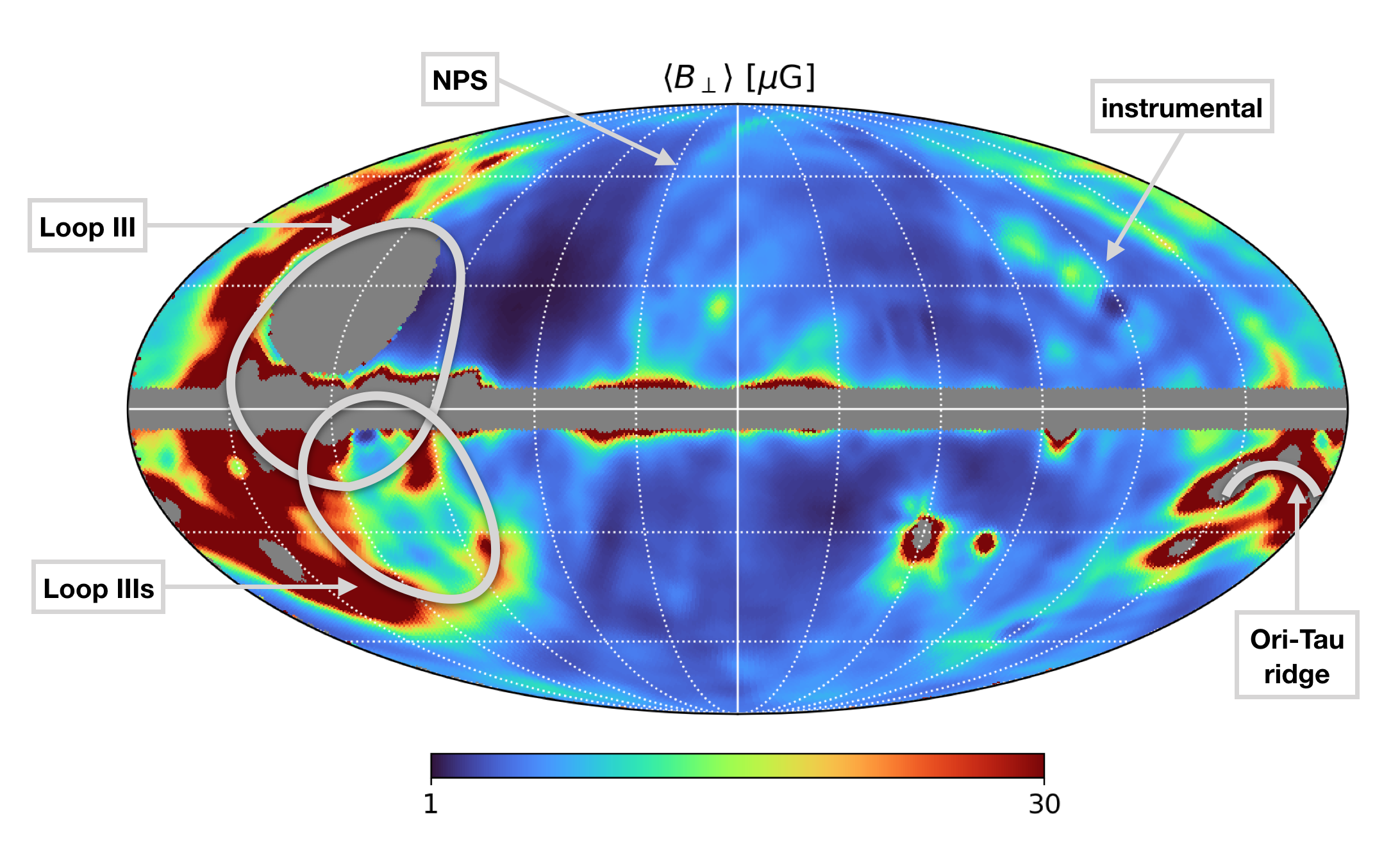}}
\caption{Map of the plane-of-the-sky magnetic-field strength averaged along the line of sight, $\langle B_{\perp} \rangle$, derived with \betaobs\ from \citetalias{Guzman2011} (see orange histogram in Fig.~\ref{fig:histoguz}) and assuming a spatially uniform $j_e$ described by model A. We label one possible instrumental pattern and a few known radio structures: the Orion-Taurus ridge \citep[Ori-Tau ridge,][]{Bracco2023}, the North Polar Spur (NPS), Loop III, and Loop IIIs \citep[e.g.,][]{vidal15}. A Galactic coordinate grid centered in the Galactic center with steps of $30^{\circ}$ in latitude and longitude is overlaid.}
\label{fig:Bmap}
\end{center}
\end{figure}
Assuming model A for $j_e$, in this section we derive a map of \bperpm\ from the \betaobs\ map of \citetalias{Guzman2011} (Sect.~\ref{ssec:Bmap}). We compare these results with synthetic data of synchrotron diffuse emission (Sect.~\ref{ssec:sims}).

\subsection{All-sky $\langle B_{\perp} \rangle$ map from \betaobs\ between 45 and 408 MHz}\label{ssec:Bmap}

In the previous sections, we have demonstrated that model A is capable of statistically reproducing the correct values of \betaobs\ observed at intermediate-to-high Galactic latitudes, both by fitting the available $j_e$ data and for values of \bperpm\ lower than 10 $\mu$G. 
With this model, we utilize the corresponding functional form of \betasyn, which is dependent on \bperpm, as depicted in red in the right panel of Fig.~\ref{fig:models}, to interpolate the \betaobs\ map presented in \citetalias{Guzman2011} and infer the respective map of \bperpm. 

The resulting all-sky map is displayed in Fig.~\ref{fig:Bmap}, shown with a linear scale between 1 $\mu$G and 30 $\mu$G. A circular region around the North Celestial Pole is masked as in \citetalias{Guzman2011}. We notice that most of the sky above $|b| > 5^{\circ}$ is consistent with values of \bperpm\ below 15 $\mu$G, as expected from Sect.~\ref{sec:je}. This suggests that, over most of the diffuse sky, considering a uniform spatial distribution of $j_e$ to describe the energy flux of CRe is a reasonable assumption. 
\begin{figure*}[!h]
\begin{center}
\resizebox{1.0\hsize}{!}{\includegraphics{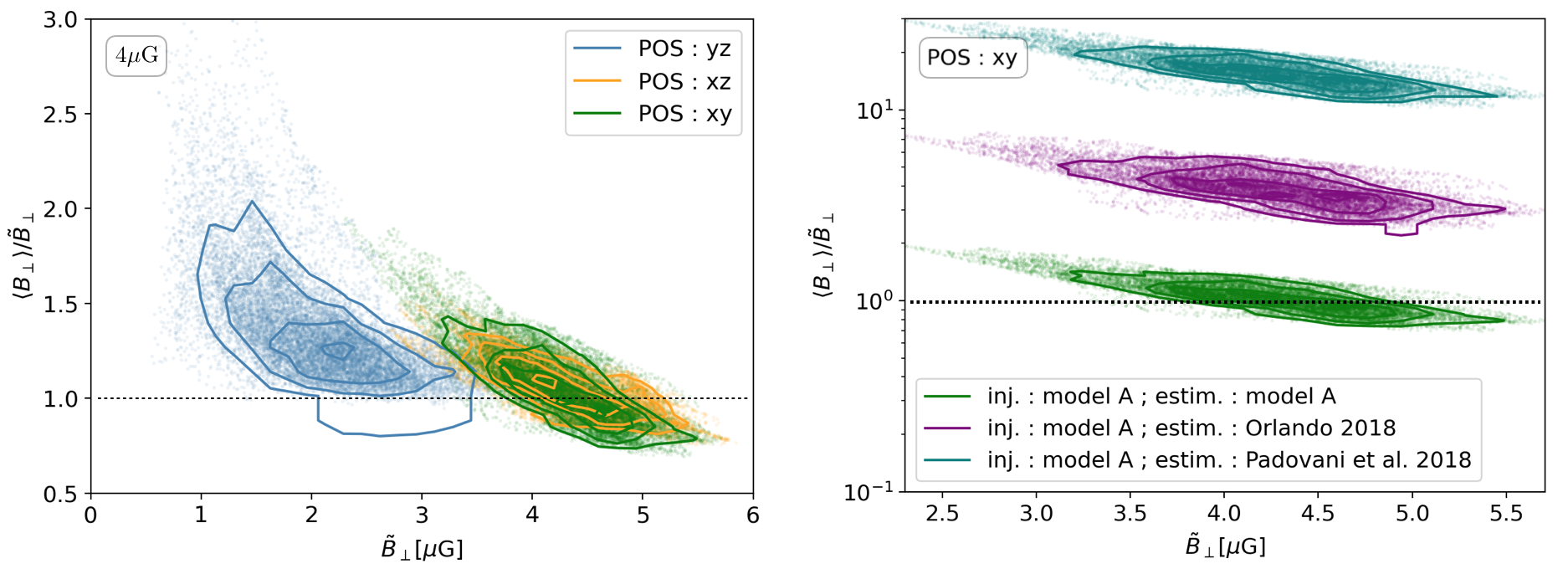}}
\caption{Estimate of $\langle B_{\perp} \rangle$ using synthetic intensity data at 45 and 408 MHz obtained with model A from one MHD simulation \citep{Bellomi2020} already implemented in \citetalias{Padovani2021} to study diffuse Galactic synchrotron emission. The initial mean magnetic-field strength of the simulation is 4 $\mu$G. The value of $\langle B_{\perp} \rangle$ is estimated as in Fig.~\ref{fig:Bmap}. {\it Left panel}: the ratio between the estimated $\langle B_{\perp} \rangle$ and the true median value of the plane-of-the-sky magnetic field strength per pixel, $\tilde{B}_{\perp}$, is shown as a function of $\tilde{B}_{\perp}$ for the three integration axes. {\it Right panel}: for the green case shown in the left panel, given the injected $j_e$-shape of model A in the simulations, $\langle B_{\perp} \rangle$ is also estimated using presently existing models, i.e. \citet{Orlando18} (in purple) and \citet{Padovani2018} (in teal). In all panels levels show the (0.5, 1, 1.5, 2)-$\sigma$ equivalent contours of the corresponding density of points.}
\label{fig:Bsims}
\end{center}
\end{figure*}

However, a few regions stand out of the map. On one hand some striped patterns in the West may be associated with instrumental artifacts present in the \citetalias{Guzman2011} map. On the other hand, most of the notable structures 
correspond to well-known arcs in the radio sky, such as the Orion-Taurus ridge \citep[Ori-Tau ridge,][]{planckXXV2016, Bracco2023}, the North Polar Spur (NPS), and Loops III-IIIs \citep[e.g.,][]{vidal15}. In Sect.~\ref{sec:discussion}, we discuss possible reasons that could explain their special appearance on the map with values of \bperpm\ significantly larger than 15 $\mu$G. Finally, some residual contamination of free-free absorption in the proximity of the Galactic plane could bias the estimate of \bperpm\ toward large values (e.g., between Galactic longitudes, $-30^{\circ} < l < +30^{\circ}$).

\subsection{Validation of the \bperpm\ map with synthetic data}\label{ssec:sims}

We validate the methodology used to derive the \bperpm\ map by using synthetic data of synchrotron emission at 45 MHz and 408 MHz obtained from one state-of-the-art MHD numerical simulation of the multiphase and magnetized ISM \citep{Bellomi2020}. In summary, this numerical model, generated using the RAMSES code \citep{Teyssier2002, Fromang2006}, simulates a 50-pc cube with 128 pixels per side of warm-diffuse gas (with initial temperatures and volume densities of $T = 8000$ K and $n_{\rm H} = 1.5$ cm$^{-3}$, respectively), where cold and dense structures ($T < 100$ K and $n_{\rm H} > 100$ cm$^{-3}$) form due to thermal instability and hydro-magnetic turbulence. The initial magnetic field considered in the simulation aligns with typical values in the diffuse ISM (4 $\mu$G) and is directed along one of the main axes of the cube (hereafter, the $x$ axis). 

This model has already been utilized in \citetalias{Padovani2021} to simulate radio emission below 200 MHz. We follow the same recipe and derive $T_{\rm b}$ by means of Eqs.~\ref{eq:epsnu} and \ref{eq:tb}, where $\bperp$ varies depending on the choice of the main axis of the cube considered for the integration of Eq.~\ref{eq:tb}, and $j_e$ is described by model A. Given the simulated $T_{\rm b}$ maps, we measure \betasyn\ between 45 MHz and 408 MHz for each LOS as done with real observations. Since we do not apply any Gaussian beam to the synthetic data, the choice of $\theta_{\rm T}$ does not affect the estimate of \betasyn. From the \betasyn\ maps, we interpolate back the value of \bperpm\ using model A (see Sect.~\ref{ssec:Bmap}) and compare it with the true median value of the simulated magnetic field along each sight line (\bperpT).

In the left panel of Fig.~\ref{fig:Bsims}, we show the ratio between \bperpm\ and \bperpT\ as a function of the latter one. We explore three cases depending on the integration axis, namely one along (plane of the sky -- POS -- $yz$) and two across (POS, $xy$ and $xz$) the mean magnetic-field direction. 

Our methodology generally results in an excess bias toward the actual value of \bperpT, which becomes more pronounced, up to a factor of 2, as the magnetic-field strength decreases. Consequently, the most severe bias arises when integrating the cube along the mean direction of the magnetic field. In the other two cases our estimates are highly consistent (within 1-$\sigma$) with the true value of the magnetic field. 

In the right panel of Fig.~\ref{fig:Bsims}, for the POS equal to the $xy$ plane, we test the bias on the estimate of \bperpT\ in the case of misinterpreting the energy flux of CRe. While always considering model A in the synthetic data, we interpolate back \bperpm\ referring to the models of \citet{Padovani2018} and \citet{Orlando18} in teal and purple colours, respectively. As expected from Fig.~\ref{fig:models}, both cases significantly overestimate \bperpT\ compared to model A, reaching excess biases of more than one order of magnitude in the case of \citet{Padovani2018}. 

\section{Discussion}\label{sec:discussion}

We have introduced a new analytical model for $j_e$ that depends on a minimal set of parameters, making it easily usable for modeling the diffuse synchrotron emission of the Galaxy. Nevertheless, it is important to clarify that our phenomenological model does not represent the injection spectrum of CRe at the emission source; instead, it corresponds to the effective CRe spectrum after propagation and attenuation processes.

We have shown that, over most of the sky, $j_e$ can be considered rather spatially uniform (at least at the angular scales probed by this work). A set of model parameters (see model A in Table~\ref{tab:models}) can be chosen such that the values of \bperpm\ needed to reproduce available \betaobs\ at low frequencies are broadly consistent with the magnetic-field strength measured in the diffuse ISM below 10 $\mu$G \citep[e.g.,][]{Heiles2005, Sobey2019}. This is displayed in the all-sky map of \bperpm\ shown in Fig.~\ref{fig:Bmap}. However, a few structures coinciding with known arcs in the radio sky seem to distinguish themselves from the average trend with anomalously large values of \bperpm. The origin of these structures is highly debated \citep[e.g.,][]{Sofue2015, Kataoka2021, Panopoulou2021}. They could represent the relic of several supernova remnants, which may affect the energy distribution of CRe in the presence of stronger and compressed magnetic fields \citep[e.g.,][]{Morlino2021}. As an example, in Fig.~\ref{fig:oriona}, we show the same as in Fig.~\ref{fig:models} but for the \betaobs\ values measured by \citet{Bracco2023} in one radio arc, namely the Ori-Tau ridge. In this case, it is not possible to reproduce the observations between $-2.3$ and $-2.2$ using model A unless resorting to \bperpm\ values of hundreds of $\mu$G. Conversely, the range of \betaobs\ in the Ori-Tau ridge can be accounted for with more standard values of \bperpm\ using our model E in Table~\ref{tab:models}. Model E adjusts the CRe population at low energy (below 1 GeV), departing from the local ISM conditions probed by the Voyager spacecraft.  
\begin{figure}[!t]
\begin{center}
\resizebox{1.0\hsize}{!}{\includegraphics{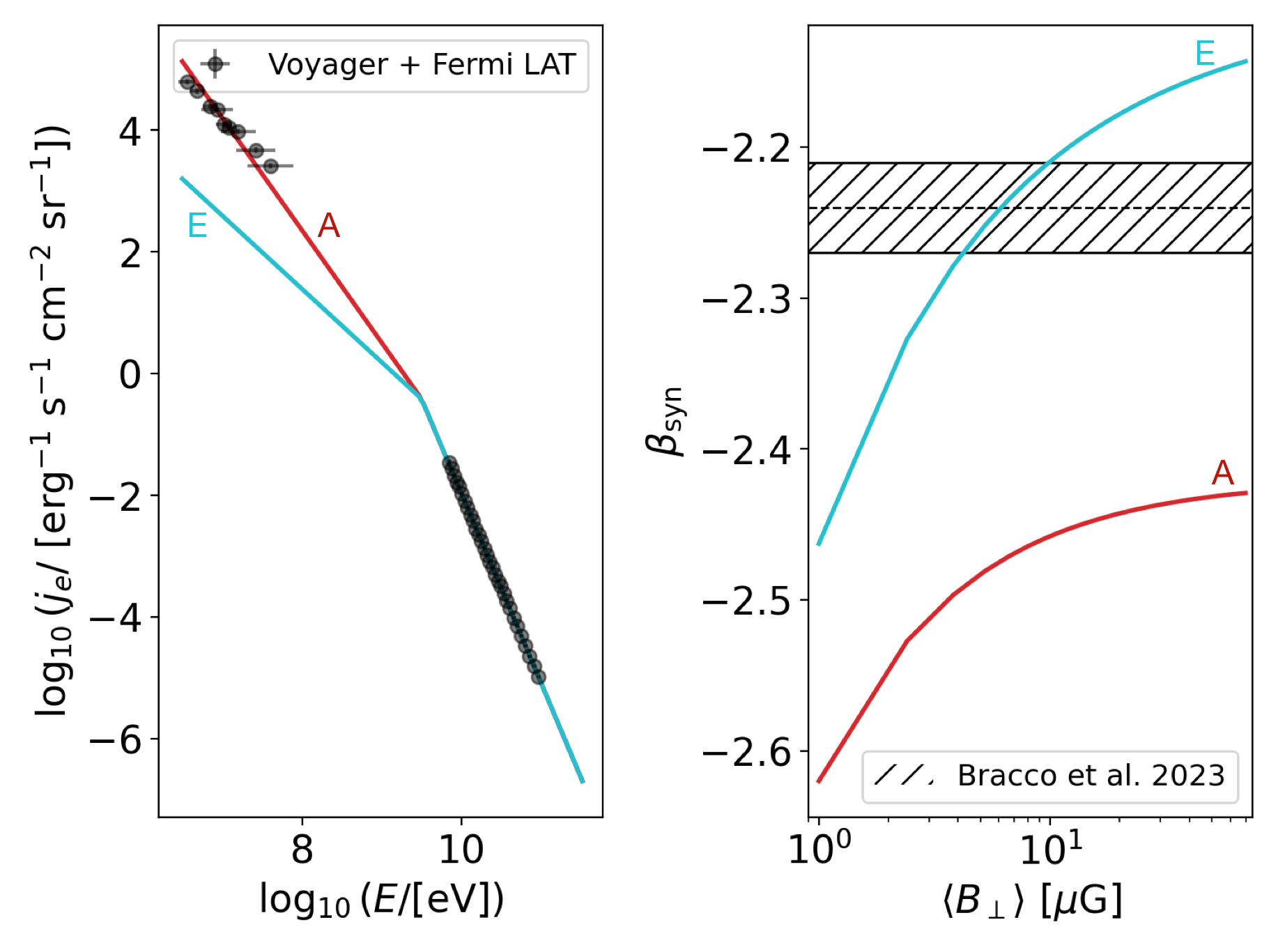}}
\caption{Same as in the left and right panels of Fig.~\ref{fig:models} but for the observed values of \betaobs\ in the Orion-Taurus ridge in hatches \citep{Bracco2023}. Models A and E for $\beta_{\rm syn}$ (right) and $j_e$ (left) are overlaid in red and cyan lines, respectively (see Table~\ref{tab:models}).}
\label{fig:oriona}
\end{center}
\end{figure}

This example illustrates the versatility of our model and its ability to be tailored to various astrophysical scenarios. Such adaptability is crucial for optimizing existing modelling of synchrotron emission \citep[e.g.,][]{Bracco2022}, especially as new discoveries at low frequencies continue to emerge. For instance, our model could help us guide the interpretation of recent and unprecedented findings made with the LOFAR telescope, one of the main SKA precursors below 200 MHz \citep[e.g.,][]{Zaroubi2015, vanEck2017, Jelic2018, VanEck2019, Bracco2020b, Turic2021}. These observations highlighted the unexpected correlation between synchrotron polarization, affected by Faraday rotation \citep{Ferriere2021}, with tracers of the neutral interstellar gas, clearing the path to a new perspective on structure formation in the multiphase and magnetized Galactic ISM.

\section{Summary}\label{sec:sum}

Cosmic-ray electrons (CRe) with energies around 1 GeV are expected to be the primary contributors to the diffuse synchrotron radiation observed in the Galaxy at radio frequencies below 1 GHz. However, determining their exact energy flux ($j_e$) has proven challenging due to solar modulation occurring between 100 MeV and 10 GeV. Previous attempts to reconcile the radio non-thermal emission with $j_e$ within this energy range have succeeded at the cost of requiring unexpectedly large values for the average Galactic magnetic-field strength perpendicular to the sight-line (\bperpm). 

In this work, we have introduced a new analytical model for $j_e$, capable of fitting ancillary data on the CRe energy flux (i.e., Voyager and Fermi LAT) while meeting observational constraints on the spectral index of the brightness temperature at low frequencies, denoted as \betaobs. The model achieves this while maintaining values of \bperpm\ consistent with existing measurements in the diffuse ISM below 10 $\mu$G \citep[e.g.,][]{Crutcher2012}. We used the \betaobs\ map obtained between 45 MHz and 408 MHz by \citetalias{Guzman2011}, found the best model that reproduces these observations, and validated its performance against various measurements of \betaobs\ in the literature between 50 MHz and 1 GHz. 

Our fiducial model (model A in Table~\ref{tab:models}) allowed us to derive, for the first time, an all-sky map of \bperpm\ that we tested against synthetic data of synchrotron emission produced from state-of-the-art MHD simulations of the multiphase and magnetized ISM. 

We found that, up to angular scales of a degree level, a uniform spatial distribution of $j_e$ can be assumed for diffuse emission at intermediate-to-high Galactic latitudes. Only a few regions in the sky, which correspond to known arc-like structures in the radio band, 
are at variance with the average trend. We demonstrated the versatility of our model to account for these anomalous regions in the case of the Orion-Taurus ridge \citep{Bracco2023}.

This work represents a useful step forward in modeling Galactic diffuse emission at low frequency given the adaptability of our data-driven four-parameter model of $j_e$. It marks an important milestone in the interpretation of current low-frequency observations of the diffuse ISM, such as those from the LOFAR telescope, and for future analyses with the SKA coming online.

\begin{acknowledgements}
We express our gratitude to the referee, Andy Strong, whose comments improved the quality of this manuscript. We are grateful to Elena Bellomi, Benjamin Godard, and Patrick Hennebelle for providing us with the MHD simulations that were used to generate synthetic data of synchrotron emission. We are thankful to Marta Spinelli for discussions on low-frequency observations of the synchrotron spectral index. AB acknowledges financial support from the INAF initiative ``IAF Astronomy Fellowships in Italy'' (grant name MEGASKAT). AB also thanks Elio for setting the pace with his memorable notes  \href{http://www.youtube.com/watch?v=wIeM4PM7aZw&ab_channel=DidierJeunesse-Descomptinesetdeshistoires}{"Ol\'el\'e moliba makasi"}. In the analysis we made use of {\tt astropy} \citep{astropy2018}, {\tt scipy} \citep{Virtanen2020}, and {\tt numpy} \citep{Harris2020}.
\end{acknowledgements}

\bibliographystyle{aa}
\bibliography{aanda.bbl}

\end{document}